\documentclass[preprint,eqsecnum,prd,aps,nofootinbib,showpacs]{revtex4}
\usepackage{amsmath,epsfig}
\begin{document}

\def\cstok#1{\leavevmode\thinspace\hbox{\vrule\vtop{\vbox{\hrule\kern1pt
\hbox{\vphantom{\tt/}\thinspace{\tt#1}\thinspace}}
\kern1pt\hrule}\vrule}\thinspace}

\preprint{NAPOLI DSF-2009/8}

\bibliographystyle{article}
\title{On the phase-integral method for the radial Dirac equation}

\author{Giampiero Esposito$^{1}$\thanks{
Electronic address: giampiero.esposito@na.infn.it},
Pietro Santorelli$^{2,1}$\thanks{
Electronic address: pietro.santorelli@na.infn.it}}

\affiliation{${ }^{1}$Istituto Nazionale di Fisica Nucleare, Sezione di
Napoli, Complesso Universitario di Monte S. Angelo, Via Cintia,
Edificio 6, 80126 Napoli, Italy \\
${ }^{2}$Dipartimento di Scienze Fisiche, Universit\`a di Napoli
Federico II, Complesso Universitario di Monte S. Angelo,
Via Cintia, Edificio 6, 80126 Napoli, Italy}

\vspace{0.4cm}
%\date{\today}

\begin{abstract}
In the application of potential models, the use of the Dirac
equation in central potentials remains of phenomenological interest. 
The associated set of
decoupled second-order ordinary differential equations is here studied
by exploiting the phase-integral technique,
following the work of Fr\"{o}man and Fr\"{o}man that provides a
powerful tool in ordinary quantum mechanics.
For various choices of the scalar and vector parts of the potential,
the phase-integral formulae are derived and discussed, jointly with
formulae for the evaluation of Stokes and anti-Stokes lines. A criterion
for choosing the base function in the phase-integral method is also
obtained, and tested numerically. The case of scalar confinement is then 
found to be more tractable.
\end{abstract}


\maketitle
\bigskip
\vspace{2cm}

\section{Introduction}
\label{s:1}

Several problems of interest in theoretical physics lead eventually to
the differential equation
\begin{equation}
\left({d^{2}\over dz^{2}}+R(z)\right)\psi(z)=0,
\label{(1.1)}
\end{equation}
where $R$ is a single-valued analytic function of the complex
variable $z$. The form of (1.1) suggests looking for solutions expressed
through a prefactor $A(z)$ and a phase $w(r)$, i.e.
\begin{equation}
\psi_{\pm}(z)=A(z)e^{\pm i w(z)}.
\label{(1.2)}
\end{equation}
The Wronskian of $\psi_{+}(z)$ and $\psi_{-}(z)$ is equal to
$-2iA^{2}{dw \over dz}$, and on the other hand the Wronskian of two
linearly independent solutions of Eq. (1.1) is a constant. Thus,
the prefactor $A(z)$ reads as const. $\times {1\over \sqrt{dw/dz}}$,
and one has \cite{Froman02}
\begin{equation}
\psi(z)={1\over \sqrt{q(z)}}e^{\pm i w(z)},
\label{(1.3)}
\end{equation}
where
\begin{equation}
w(z)=\int^{z}q(\zeta)d\zeta,
\label{(1.4)}
\end{equation}
the function $w$ being the {\it phase integral}, while $q$ is called the
{\it phase integrand}. Moreover, upon insertion of the exact solution (1.3),
(1.4) into Eq. (1.1), one finds that the phase integrand $q(z)$
should satisfy the $q$-equation
\begin{equation}
f(z,q(z),R(z)) \equiv
q^{-{3\over 2}}{d^{2}\over dz^{2}}q^{-{1\over 2}}
+{R(z)\over q^{2}}-1=0.
\label{(1.5)}
\end{equation}
In practice, however, the task of finding exact solutions of Eq. (1.5)
is rather difficult. The best one can do is often to determine a
function $Q$ that is an approximate solution of the $q$-equation
(1.5), so that
\begin{equation}
\varepsilon_{0} \equiv f(z,Q(z),R(z)) <<1.
\label{(1.6)}
\end{equation}
The approximate phase-integral method consists in finding approximate
solutions of Eq. (1.1) with {\it unspecified base function} $Q$. A
criterion for finding $Q$ is that the function $\varepsilon_{0}$ defined
in (1.6) should be much smaller than unity in the region of the
complex-$z$ plane relevant for the problem. However, this criterion does
not determine the base function $Q$ uniquely, the physicist has 
a whole set of basis functions $Q$ at his disposal, 
and this arbitrariness can be exploited.

On the other hand, along the years,
many efforts have been devoted in the literature to
the theoretical investigation of light fermions confined by a potential
field \cite{PHRVA-D51-5079}. In the phenomenological applications, when
dealing with mesons consisting of a heavy quark and a light quark, one can
imagine that the heavy quark is indeed very heavy and acts as a ``classical''
source that can be represented as a superposition of Coulomb-like plus 
linear potential, better known as Cornell potential \cite{Cornell}. 
The mass occurring in the Dirac equation is therefore
the mass of the light quark.
It is by now well known that, on using the
Dirac equation, only Lorentz scalar confinement leads to normalizable
stationary states, while in a suitable variant of the Dirac equation,
called ``no pair'', only Lorentz vector confinement has normal Regge
behaviour. Hereafter we focus on the stationary Dirac equation for a
quark of mass $m$ in a Lorentz scalar potential $V_{S}(r)$ and in the
time component of a Lorentz vector potential $V_{V}(r)$, i.e.
\cite{PHSTB-77-065005}
\begin{equation}
{dF \over dr}=-{\kappa \over r}F
+{mc^{2}+E+V_{S}-V_{V} \over {\hbar}c}G,
\label{(1.7)}
\end{equation}
\begin{equation}
{dG \over dr}={\kappa \over r}G
+{mc^{2}-E+V_{S}+V_{V} \over {\hbar}c}F,
\label{(1.8)}
\end{equation}
where $\kappa=-l-1$ if $j=l+{1\over 2}, \; \kappa=l$
if $j=l-{1\over 2}$. In the resulting second-order equations,
first derivatives can be removed by putting
\begin{equation}
\begin{pmatrix}
F(r) \\
G(r)
\end{pmatrix} =
\begin{pmatrix}
\sqrt{E+mc^{2}+V_{S}-V_{V}} \hfill & 0 \\
\; \; \; \; \; \; \;  0 & \sqrt{E-mc^{2}-V_{S}-V_{V}}
\end{pmatrix}
\begin{pmatrix}
f(r) \\
g(r)
\end{pmatrix}.
\label{(1.9)}
\end{equation}

Section II studies the second-order equations resulting from the radial
Dirac equations (1.7) and (1.8), preparing the ground for the application
of the phase-integral method. Section III describes various possible
choices of basis function in the phase-integral method. Sec. IV arrives
at a general criterion for choosing a suitable basis function $Q$.
Sec. V performs a numerical analysis of the applicability
of such a criterion.
Sec. VI studies Stokes and anti-Stokes lines for the squared Dirac
equation in a central potential, inspired by the choice of $Q$ made in
the simpler analysis of central potentials in ordinary quantum mechanics.
Concluding remarks and open problems are presented in Sec. VII.

\section{Second-order equations from the Radial Dirac equation}
\label{s:2}

With the notation in the Introduction, our starting point is the following
set of decoupled second-order equations obtained from the radial
Dirac equation:
\begin{equation}
\begin{pmatrix}
\frac{d^2}{dr^2} +R_{f}(r) \hfill & 0  \\
\; \; \; \; \; \; \; 0 \hfill & \frac{d^2}{dr^2}+R_{g}(r) \hfill
\end{pmatrix}
\begin{pmatrix}
f(r) \\
g(r) \hfill
\end{pmatrix}
=0,
\label{(2.1)}
\end{equation}
where the ``potential'' terms read as \cite{PHSTB-77-065005}
\begin{eqnarray}
R_{f}(r) & \equiv & {(E-V_{V})^{2}-(mc^{2}+V_{S})^{2}
\over ({\hbar c})^{2}}-{\kappa (\kappa+1)\over r^{2}} \nonumber \\
&+& \left[V_{V}'-V_{S}' \over mc^{2}+E +V_{S}-V_{V} \right]
{\kappa \over r}-{(V_{V}''-V_{S}'')\over 2[mc^{2}+E
+V_{S}-V_{V}]} \nonumber \\
&-& {3\over 4} \left[V_{V}'-V_{S}' \over
mc^{2}+E+V_{S}-V_{V} \right]^{2},
\label{(2.2)}
\end{eqnarray}
\begin{eqnarray}
R_{g}(r) & \equiv & {(E-V_{V})^{2}-(mc^{2}+V_{S})^{2}
\over ({\hbar c})^{2}}-{\kappa (\kappa-1)\over r^{2}} \nonumber \\
&+& \left[V_{V}'+V_{S}' \over mc^{2}-E +V_{S}+V_{V} \right]
{\kappa \over r}-{(V_{V}''+V_{S}'')\over 2[mc^{2}-E
+V_{S}+V_{V}]} \nonumber \\
&-& {3\over 4} \left[V_{V}'+V_{S}' \over
mc^{2}-E+V_{S}+V_{V} \right]^{2}. 
\label{(2.3)}
\end{eqnarray}
Note that, for energies $E < -mc^{2}$, the following difficulty 
arises: the relation between $f(r)$ and $F(r)$ in (1.9) becomes
singular at the point $r=r_{f}$ such that 
$V_{V}(r_{f})-V_{S}(r_{f})=E+mc^{2}$. Thus, the effective potential
$R_{f}(r)$ in (2.2) becomes infinite at $r \rightarrow r_{f}$. The
solutions become meaningless near the point $r=r_{f}$ because the
phase integrals diverge. Similar remarks 
\cite{Popov,Zeldovich,Lazur} hold for $g(r)$ and the
effective potential in (2.3). However, this difficulty is purely
formal because the original Dirac system (1.7) and (1.8) is not
singular at the point $r=r_{f}$. A powerful JWKB analysis of the
first-order Dirac system (1.7) and (1.8) 
can be found in \cite{Lazur}.  

Equations (2.1)--(2.3) suggest exploiting the known properties of
the differential equation (1.1), which, as we said,
is much studied in classical mathematical physics and ordinary
quantum mechanics. The change of dependent and independent variable
that preserves the form of (1.1) without first derivative is given by
\begin{equation}
\psi(z) \equiv {1\over \sqrt{Q(z)}}\varphi(z),
\label{(2.4)}
\end{equation}
\begin{equation}
w(z) \equiv \int^{z}Q(\zeta)d\zeta,
\label{(2.5)}
\end{equation}
where the function $Q$ is not specified for the time being but will be
suitably chosen later.

Upon defining
\begin{equation}
\varepsilon \equiv {R\over Q^{2}}-1+Q^{-3/2}{d^{2}\over dz^{2}}
(Q^{-1/2}),
\label{(2.6)}
\end{equation}
equation (1.1) can be expressed in the equivalent form
\begin{equation}
\left[{d^{2}\over dw^{2}}+(1+\varepsilon)\right]\varphi(w)=0.
\label{(2.7)}
\end{equation}
Equation (2.7) is more convenient because it can be turned into a system
of two linear differential equations of the first order. For this purpose,
one assumes that the complex $w$-plane is cut in such a way that the
functions appearing are all single-valued and $\varphi$ can read as
\begin{equation}
\varphi(w)=a_{1}(w)e^{iw}+a_{2}(w)e^{-iw}.
\label{(2.8)}
\end{equation}
If we further impose that
\begin{equation}
a_{1}'(w)e^{iw}+a_{2}'(w)e^{-iw}=0,
\label{(2.9)}
\end{equation}
the first derivative of $\varphi$ reduces to
\begin{equation}
{d \varphi \over dw}=ia_{1}e^{iw}-ia_{2}e^{-iw},
\label{(2.10)}
\end{equation}
and one obtains the desired system of two first-order ordinary
differential equations, i.e. \cite{Froman65}
\begin{equation}
{da_{1}\over dw}={i\over 2}\varepsilon \Bigr(a_{1}
+a_{2}e^{-2iw}\Bigr),
\label{(2.11)}
\end{equation}
\begin{equation}
{da_{2}\over dw}=-{i\over 2}\varepsilon \Bigr(a_{2}
+a_{1}e^{2iw}\Bigr).
\label{(2.12)}
\end{equation}
Such a system can be written in matrix form as
\begin{equation}
{da \over dw}=M(w)a,
\label{(2.13)}
\end{equation}
having set
\begin{equation}
M(w) \equiv {i\over 2}\varepsilon
\begin{pmatrix}
1 \hfill & e^{-2iw} \\
-e^{2iw} \hfill & -1 \hfill
\end{pmatrix},
\label{(2.14)}
\end{equation}
\begin{equation}
a(w)=
\begin{pmatrix}
a_{1}(w) \\
a_{2}(w) \hfill
\end{pmatrix}.
\label{(2.15)}
\end{equation}
At this stage, one can replace the differential equation (2.13)
by the integral equation
\begin{equation}
a(w)=a(w_{0})+\int_{w_{0}}^{w}M(w_{1}) a(w_{1}) dw_{1},
\label{(2.16)}
\end{equation}
which can be solved by iteration, starting from the solution formula
\begin{equation}
a(w)=F(w,w_{0})a(w_{0}),
\label{(2.17)}
\end{equation}
where
\begin{eqnarray}
F(w,w_{0})&=& 1+\int_{w_{0}}^{w}dw_{1}M(w_{1})
+\int_{w_{0}}^{w}dw_{1}M(w_{1})
\int_{w_{0}}^{w_{1}}dw_{2}M(w_{2}) \nonumber \\
&+& \int_{w_{0}}^{w}dw_{1}M(w_{1})
\int_{w_{0}}^{w_{1}}dw_{2}M(w_{2})
\int_{w_{0}}^{w_{2}}dw_{3}M(w_{3})
+ ...
\label{(2.18)}
\end{eqnarray}
Under the assumption that
\begin{equation}
\sum_{j}|M_{ij}(w)| \leq m(w),
\label{(2.19)}
\end{equation}
where $m(w)$ is a non-negative quantity, one finds that, in any region
of the complex-$w$ plane where the integral
$\int_{w_{0}}^{w}m(w_{1})dw_{1}$ is bounded, the series in (2.18)
is absolutely and uniformly convergent. From (2.4), the original
equation (1.1) is then solved by
\begin{equation}
\psi(z)=a_{1}{e^{iw(z)}\over \sqrt{Q(z)}}
+a_{2}{e^{-iw(z)}\over \sqrt{Q(z)}}
=a_{1}(z)f_{1}(z)+a_{2}(z)f_{2}(z),
\label{(2.20)}
\end{equation}
where
\begin{equation}
f_{1}(z) \equiv {1\over \sqrt{Q(z)}}e^{iw(z)}, \;
f_{2}(z) \equiv {1\over \sqrt{Q(z)}}e^{-iw(z)}.
\label{(2.21)}
\end{equation}
Our main source on this topic,
ref. \cite{Froman65}, contains all details about useful approximate
formulae for the $F$-matrix and many peculiar properties of the
phase-integral approximation, which should not be confused with the
JWKB method \cite{Froman02}.

\section{Choice of the base function}
\label{s:3}

The function $Q$ in Sec. II need not coincide, when squared up,
with the function $R$ in Eq. (1.1). A
guiding principle in the choice of base function is as follows: first
find the pole of higher order (if any) in $R(z)$, and then choose
$Q(z)$ in such a way that it cancels exactly such a pole (see below).

\subsection{Scalar confinement}
\label{s:casoA}

For example, the scalar confinement is achieved with the potentials
\cite{PHRVA-D51-5079}
\begin{equation}
V_{S}=ar, \; V_{V}=0,
\label{(3.1)}
\end{equation}
for which the ``potential terms'' $R_{f}$ and $R_{g}$ in (2.2) and
(2.3) reduce to
\begin{equation}
R_{f}= {E^{2}-(mc^{2}+ar)^{2} \over ({\hbar}c)^{2}}
-{\kappa (\kappa+1)\over r^{2}}
-{a\over (mc^{2}+E+ar)}{\kappa \over r}
-{3\over 4}{a^{2}\over (mc^{2}+E+ar)^{2}},
\label{(3.2)}
\end{equation}
\begin{equation}
R_{g}= {E^{2}-(mc^{2}+ar)^{2} \over ({\hbar}c)^{2}}
-{\kappa (\kappa-1)\over r^{2}}
+{a\over (mc^{2}-E+ar)}{\kappa \over r}
-{3\over 4}{a^{2}\over (mc^{2}-E+ar)^{2}}.
\label{(3.3)}
\end{equation}
The experience gained in ordinary quantum mechanics suggests
therefore choosing \cite{Froman65}
\begin{equation}
Q_{f}^{2} \equiv R_{f}+{\kappa (\kappa+1)\over r^{2}},
\label{(3.4)}
\end{equation}
\begin{equation}
Q_{g}^{2} \equiv R_{g}+{\kappa (\kappa-1)\over r^{2}}.
\label{(3.5)}
\end{equation}

\subsection{Logarithmic potential}
\label{s:casoB}

More generally, however, bearing in mind that singularities in
(2.2) and (2.3) might receive a further contribution from $V_{V}$
or $V_{S}$ if they were of logarithmic type, one can take
\begin{equation}
V_{S}={1\over a}\log \left({r\over r_{0}}\right), \;
V_{V}=0,
\label{(3.6)}
\end{equation}
bearing also in mind that only a scalar potential is able to confine a
quark in the Dirac equation, and that a relativistic
$Q{\overline q}$ system is indeed well described by the choice (3.6),
as shown in \cite{PHLTA-B97-143}. The potential terms $R_{f}$ and
$R_{g}$ in (2.2) and (2.3) are then found to develop also a
logarithmic singularity at $r=0$, because the l'Hospital rule for
taking limits implies that
$$
\lim_{r \to 0}{1\over r^{2} \log(r)}
=\lim_{r \to 0}{1\over r^{2}\log^{2}(r)}=\infty.
$$
We are then led to get rid of both the pole-like and logarithmic
singularities of $Q$ at $r=0$, by defining
\begin{equation}
Q_{f}^{2} \equiv R_{f}+{\kappa(\kappa+1)\over r^{2}}
+{\left(\kappa+{1\over 2}\right)\over r^{2}\log (r/r_{0})}
+{3\over 4r^{2}} {1\over \log^{2} (r/r_{0})},
\label{(3.7)}
\end{equation}
\begin{equation}
Q_{g}^{2} \equiv R_{g}+{\kappa(\kappa-1)\over r^{2}}
-{\left(\kappa+{1\over 2}\right)\over r^{2}\log (r/r_{0})}
+{3\over 4r^{2}} {1\over \log^{2} (r/r_{0})}.
\label{(3.8)}
\end{equation}
Interestingly, we are suggesting a novel perspective on the
logarithmic potential, arriving at it from the point of view of the
singularity structure of the base function in the phase-integral method.

\subsection{A linear plus Coulomb-type potential}
\label{s:casoC}

One can also consider the Cornell potential \cite{Cornell} which is linear in the scalar
part and of Coulomb-type in the vector part, i.e.
\begin{equation}
V_{S}=ar, \; V_{V}={b\over r}.
\label{(3.9)}
\end{equation}
As $r \rightarrow 0$, the centrifugal term
${\kappa(\kappa \pm 1)\over r^{2}}$ in $R_{f}$ (respectively $R_{g}$)
is then found to receive further contributions with a second-order
pole at the origin, so that we can remove such a singularity in $Q$
by defining
\begin{equation*}
Q_{f}^{2} \equiv R_{f}+{\left[\kappa^{2}-{1\over 4}
-(b / {\hbar} c)^{2}\right] \over r^{2}},
\end{equation*}
\begin{equation*}
Q_{g}^{2} \equiv R_{g}+{\left[\kappa^{2}+{7\over 4}
-(b / {\hbar} c)^{2}\right] \over r^{2}}.
\end{equation*}
However, the resulting integral (2.5) for the independent variable $w$ is
too complicated for analytic or numerical purposes.

\subsection{Analogy with central potentials in ordinary quantum mechanics}
\label{s:casoD}

It is therefore more convenient, in our relativistic problem,
to fully exploit the arbitrariness of the base
function $Q$ by defining it in such a way that it coincides with the
form taken by $Q$ in non-relativistic problems in a central potential. For
example, for the Schr\"{o}dinger equation in a central potential it is
helpful to deal with a $Q$ function of the form \cite{Froman65}
$Q^{2}=1+{2 \eta \over r}$. In our problem, both $R_{f}$ in (2.2) and
$R_{g}$ in (2.3) contain exactly, i.e. without making any expansion,
the term $-{2Eb \over ({\hbar}c)^{2}}{1\over r}$, which is indeed of the
form ${2 \eta \over r}$ with
\begin{equation}
\eta \equiv -{Eb \over ({\hbar}c)^{2}}.
\label{(3.10)}
\end{equation}
We thus look for
\begin{equation}
Q_{f}^{2}(r)=R_{f}(r)+u_{f}(r)=1+{2 \eta \over r}.
\label{(3.11)}
\end{equation}
In this equation, the desired additional term can be obtained in exact
form as
$$
u_{f}(r)=1+{2 \eta \over r}-R_{f}(r),
$$
where $R_{f}$ leads to exact cancellation of the terms proportional to
${1\over r}$. We then find, from (2.5) and (3.11) (see
\cite{Froman65}),
\begin{equation}
w_{f}(r)=2\eta \left \{ \sqrt{{r\over 2\eta}\left(1
+{r\over 2\eta}\right)}
+\log \left[\sqrt{{r\over 2\eta}}
+\sqrt{1+{r\over 2\eta}}\right]\right \},
\label{(3.12)}
\end{equation}
and, from (2.6),
\begin{equation}
\varepsilon_{f}={R_{f}\over Q_{f}^{2}}-1+Q_{f}^{-{3\over 2}}
{d^{2}\over dr^{2}} Q_{f}^{-{1\over 2}}
={R_{f}\over Q_{f}^{2}}-1-Q_{f}^{-{1\over 2}}
{d^{2}\over dw^{2}}Q_{f}^{1\over 2},
\label{(3.13)}
\end{equation}
which yield, by virtue of (2.21),
\begin{equation}
f(r)=a_{1,f}{e^{i w_{f}(r)}\over \sqrt{Q_{f}(r)}}
+a_{2,f}{e^{-i w_{f}(r)}\over \sqrt{Q_{f}(r)}},
\label{(3.14)}
\end{equation}
where the functions $a_{1,f}$ and $a_{2,f}$ can be obtained from
(2.13)--(2.18), with $\varepsilon=\varepsilon_{f}$ in (2.14).

By following an analogous procedure, we find
\begin{equation}
Q_{g}^{2}(r)=R_{g}(r)+u_{g}(r)=1+{2 \eta \over r}=Q_{f}^{2}(r),
\label{(3.15)}
\end{equation}
\begin{equation}
w_{g}(r)=w_{f}(r),
\label{(3.16)}
\end{equation}
\begin{equation}
\varepsilon_{g}={R_{g}\over Q_{g}^{2}}-1+Q_{g}^{-{3\over 2}}
{d^{2}\over dr^{2}}Q_{g}^{-{1\over 2}}
={R_{g}\over Q_{g}^{2}}-1-Q_{g}^{-{1\over 2}}
{d^{2}\over dw^{2}}Q_{g}^{1\over 2},
\label{(3.17)}
\end{equation}
\begin{equation}
g(r)=a_{1,g}{e^{iw_{g}(r)}\over \sqrt{Q_{g}(r)}}
+a_{2,g}{e^{-i w_{g}(r)}\over \sqrt{Q_{g}(r)}},
\label{(3.18)}
\end{equation}
bearing in mind that $R_{g} \not = R_{f} \Longrightarrow
\varepsilon_{g} \not = \varepsilon_{f}$, and setting now
$\varepsilon=\varepsilon_{g}$ in (2.14) for the evaluation of
$a_{1,g}$ and $a_{2,g}$.

We should now recall that, by virtue of the identity \cite{Froman65}
\begin{eqnarray}
\; & \; & M(w_{1})M(w_{2})...M(w_{n}) \nonumber \\
&=& \left({i\over 2}\right)^{n} \varepsilon(w_{1})
\varepsilon(w_{2})... \varepsilon(w_{n})
\left[1-e^{-2i(w_{1}-w_{2})}\right]
\left[1-e^{-2i(w_{2}-w_{3})}\right] \nonumber \\
&...& \left[1-e^{-2i(w_{n-1}-w_{n})}\right]
\begin{pmatrix}
1 \hfill & e^{-2iw_{n}} \\
-e^{2iw_{1}} \hfill & -e^{2i(w_{1}-w_{n})}
\end{pmatrix},
\label{(3.19)}
\end{eqnarray}
the $F$-matrix in (2.17)-(2.18) can be expressed through a fairly
simple series, i.e. \cite{Froman65}
\begin{eqnarray}
\; & \; & F_{11}(w,w_{0})=1+\int_{w_{0}}^{w}dw_{1} \; {i\over 2}
\varepsilon(w_{1}) \nonumber \\
&+& \int_{w_{0}}^{w}dw_{1}\; {i\over 2}\varepsilon(w_{1})
\int_{w_{0}}^{w_{1}}dw_{2}\; {i\over 2}\varepsilon(w_{2})
\left[1-e^{-2i(w_{1}-w_{2})}\right] \nonumber \\
&+& ...,
\label{(3.20)}
\end{eqnarray}
\begin{eqnarray}
\; & \; & F_{12}(w,w_{0})=\int_{w_{0}}^{w}dw_{1} \; {i\over 2}
\varepsilon(w_{1}) e^{-2iw_{1}} \nonumber \\
&+& \int_{w_{0}}^{w}dw_{1}\; {i\over 2}\varepsilon(w_{1})
\int_{w_{0}}^{w_{1}}dw_{2}\; {i\over 2}\varepsilon(w_{2})
\left[1-e^{-2i(w_{1}-w_{2})}\right]e^{-2iw_{2}} \nonumber \\
&+& ...,
\label{(3.21)}
\end{eqnarray}
\begin{eqnarray}
\; & \; & F_{21}(w,w_{0})=-\int_{w_{0}}^{w}dw_{1} \; {i\over 2}
\varepsilon(w_{1}) e^{2iw_{1}} \nonumber \\
&-& \int_{w_{0}}^{w}dw_{1}\; {i\over 2}\varepsilon(w_{1})
e^{2iw_{1}}
\int_{w_{0}}^{w_{1}}dw_{2}\; {i\over 2}\varepsilon(w_{2})
\left[1-e^{-2i(w_{1}-w_{2})}\right] \nonumber \\
&+& ...,
\label{(3.22)}
\end{eqnarray}
\begin{eqnarray}
\; & \; & F_{22}(w,w_{0})=1-\int_{w_{0}}^{w}dw_{1} \; {i\over 2}
\varepsilon(w_{1}) \nonumber \\
&-& \int_{w_{0}}^{w}dw_{1}\; {i\over 2}\varepsilon(w_{1})
\int_{w_{0}}^{w_{1}}dw_{2}\; {i\over 2}\varepsilon(w_{2})
\left[1-e^{-2i(w_{1}-w_{2})}\right]e^{2i(w_{1}-w_{2})} \nonumber \\
&+& ... \; .
\label{(3.23)}
\end{eqnarray}

\section{A general criterion for choosing the base function}
\label{s:4}

We have also tried to find a base function Q by {\it assuming} its
behaviour for small and large values of r, \textit{i.e.}
\begin{equation}
Q(r) = \frac{\alpha_{1} }{r} + \alpha_{2} + \alpha_{3} r\,.
\label{(4.1)}
\end{equation}
This base function can be analytically integrated, thus, in
principle, we can obtain the phase integral according to (2.5). To fix
the free parameter entering the previous expression we assume that
the $\varepsilon$ parameter in (2.6) should
vanish at small and large distances. However, this criterion does not
ensure that $\varepsilon$ remains small throughout the whole range of
values of $r$, and we have instead found regions where the resulting
$\varepsilon$ is, regrettably, larger than $1$, thus making our choices
unsuitable. A general method is instead as follows. Since we have to
fulfill the condition (1.6) with $\varepsilon$ defined as in (2.6) and
$R=R_{f}$ or $R_{g}$, we re-express (1.6) in the form
\begin{equation}
\left | R-Q^{2}+Q^{1/2}{d^{2}\over dz^{2}}(Q^{-1/2}) \right |
<< Q^{2},
\label{(4.2)}
\end{equation}
and define
\begin{equation}
{\cal A} \equiv R-Q^{2},
\label{(4.3)}
\end{equation}
\begin{equation}
{\cal B} \equiv Q^{1/2}{d^{2}\over dz^{2}}(Q^{-1/2}),
\label{(4.4)}
\end{equation}
or, the other way around,
\begin{equation}
{\cal A} \equiv Q^{1/2} {d^{2}\over dz^{2}}(Q^{-1/2}),
\label{(4.5)}
\end{equation}
\begin{equation}
{\cal B} \equiv R - Q^{2},
\label{(4.6)}
\end{equation}
bearing in mind that
\begin{equation}
\left |\! \frac{}{}|{\cal A}|-|{\cal B}|\right | \leq |{\cal A}+{\cal B}| \leq |{\cal A}|+|{\cal B}|.
\label{(4.7)}
\end{equation}
Moreover, we can always make the conventional choice according to
which $|{\cal A}| > |{\cal B}|$.

When (4.3) and (4.4) hold, if both ${\cal A}$ and ${\cal B}$ are positive, the
conditions (1.6) and (2.6) yield
\begin{equation}
R-Q^{2}+Q^{1/2}{d^{2}\over dz^{2}}(Q^{-1/2}) \leq Q^{2},
\label{(4.8)}
\end{equation}
i.e.
\begin{equation}
R -2Q^{2}+ Q^{1/2}{d^{2}\over dz^{2}}(Q^{-1/2}) \leq 0.
\label{(4.9)}
\end{equation}
When (4.3) and (4.4) hold, if ${\cal A}>0$ and ${\cal B}<0$, conditions (1.6) and
(2.6) yield
\begin{equation}
|R-Q^{2}| - \left| Q^{1/2}{d^{2}\over dz^{2}}
(Q^{-1/2}) \right | \leq Q^{2},
\label{(4.10)}
\end{equation}
which coincides with (4.9) because ${\cal A}=R-Q^{2} >0$ while
${\cal B}=-|{\cal B}| <0$.

Nothing changes if instead (4.5) and (4.6) hold. For example, if ${\cal A}$
defined in (4.5) is positive and ${\cal  B}$ defined in (4.6) is negative,
one finds from (1.6) and (2.6)
\begin{equation}
Q^{1/2}{d^{2}\over dz^{2}}(Q^{-1/2})-|R-Q^{2}| \leq Q^{2},
\label{(4.11)}
\end{equation}
which coincides with (4.9). Thus, in all possible cases, the family
of as yet unknown base functions $Q$ has to be chosen in such a way
that the majorization (4.9) is always satisfied.

\section{Numerical results on $Q_{f,g}^{2}$}
\label{s:5}

In this section we collect all numerical results regarding the choice
of the squared base function $Q_{f,g}^{2}$ by following the
considerations in the previous sections.
First of all we work in the natural unit system ($\hbar = c = 1)$,
and we plot in the figures 1--4 the left-hand side of eq. (\ref{(4.9)}).
The chosen range for $r$ is the typical one for the heavy mesons 
phenomenology.  The numerical values for the parameter are taken from the
phenomenological analysis
of the meson spectrum by using the Dirac equation
\cite{PHRVA-D51-5079}. In particular,
we restrict ourselves to consider the numerical parameter for the
charmed particles.
Moreover, it should be observed that in \cite{PHRVA-D51-5079} only the Cornell
potential has been considered (cf. subsection \ref{s:casoC}).
However, we use the same
numerical values for parameters also in the case \ref{s:casoA},
\ref{s:casoC} and
\ref{s:casoD} because the qualitative behaviour
of the results does not depend strongly on the numerical
values of the parameters.

\begin{figure}
\begin{center}
\begin{tabular}{ccc}
\includegraphics[width=7cm]{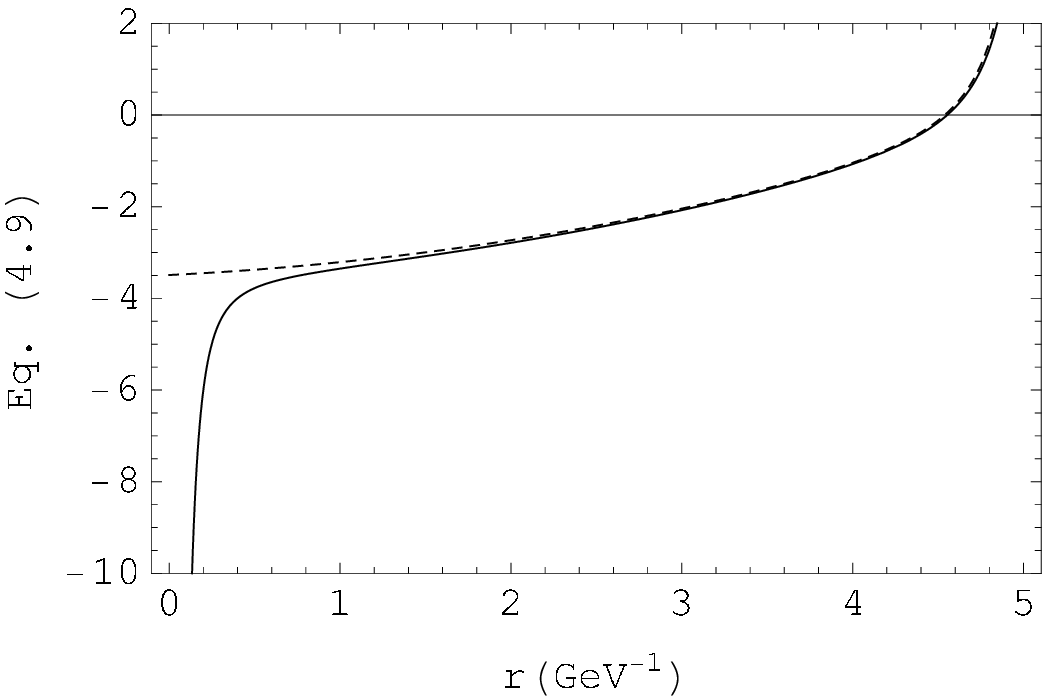} &
\hspace{2truecm}                      &
\includegraphics[width=7cm]{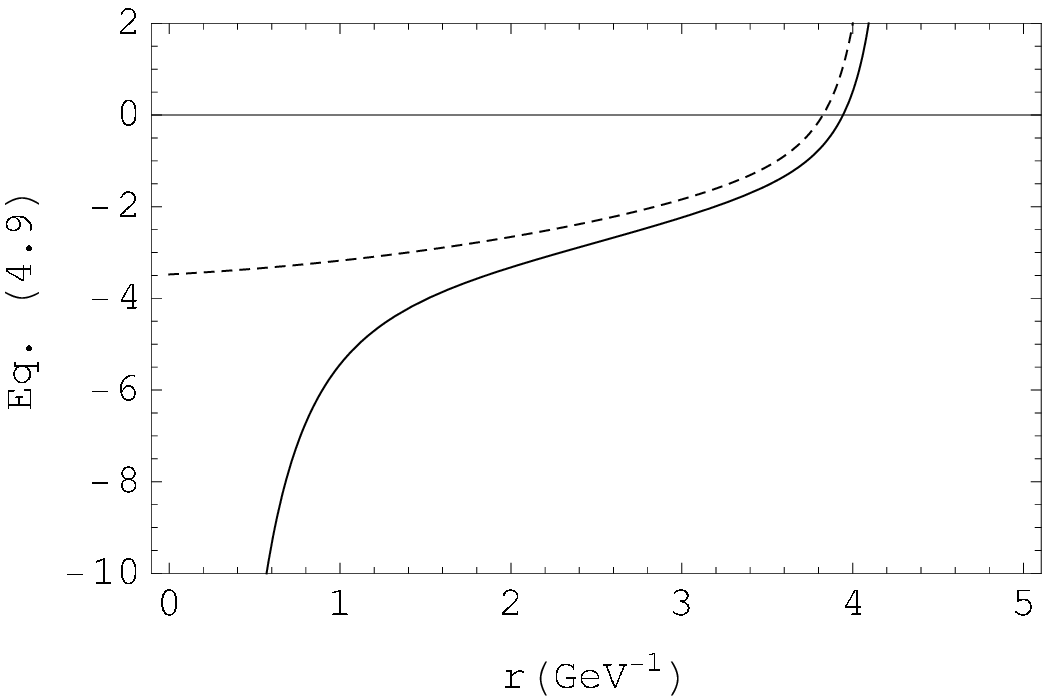}
\end{tabular}
\caption{
The left-hand side of eq. (\ref{(4.9)}) is plotted versus $r$ for Sec. IIIA.
The continuous (dashed) line corresponds to $k=-1$ ($k=0$).
We have used $m=0.300\,\,GeV$, $a=0.308\,\, GeV^{2}$ and $E = 1.9\,\,GeV$.
}
\label{f:casoA}
\end{center}
\end{figure}

\begin{figure}
\begin{center}
\begin{tabular}{ccc}
\includegraphics[width=7cm]{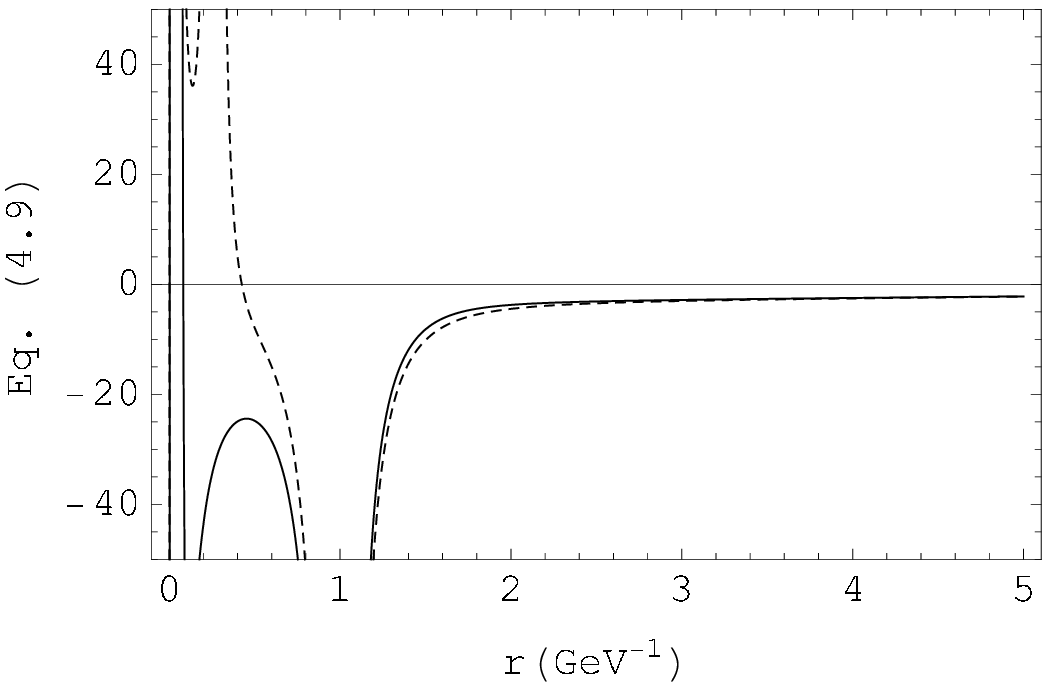} &
\hspace{2truecm}                      &
\includegraphics[width=7cm]{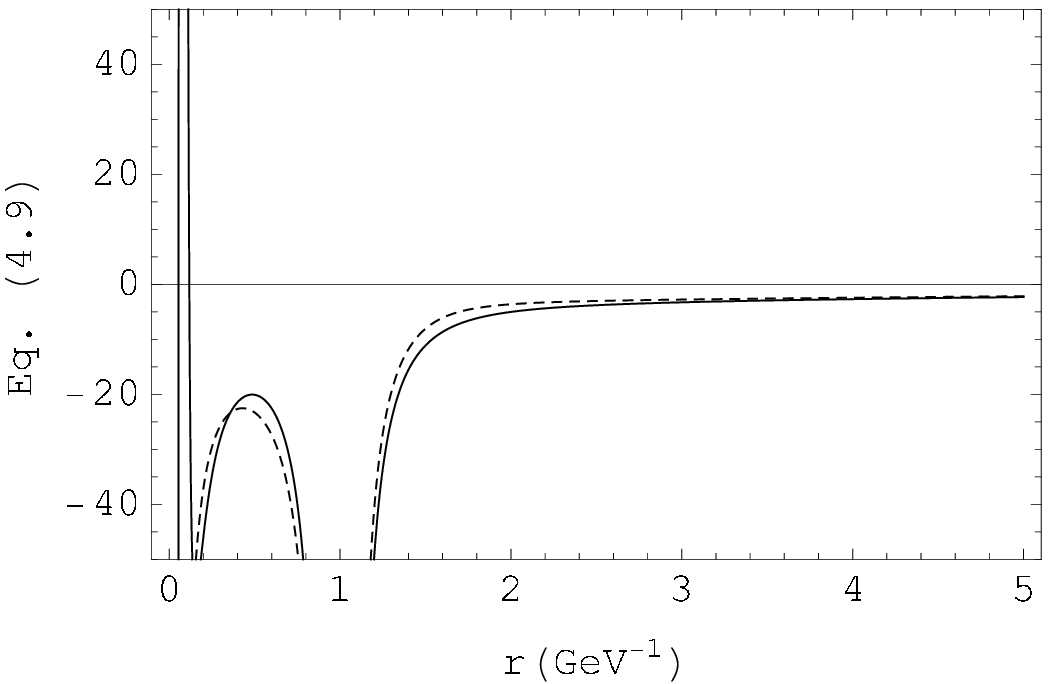}
\end{tabular}
\caption{
The left-hand side of eq. (\ref{(4.9)}) is plotted versus $r$ for Sec. IIIB.
The continuous (dashed) line corresponds to $k=-1$ ($k=0$). We have used
$m=0.300\,\,GeV$, $a=1/\sqrt{0.308}\,\, GeV^{-1}$, $E = 1.9\,\,GeV$
and $r_0= 1\,\, GeV$. The left (right) panel corresponds to the case of
$f$ ($g$) in eq. (\ref{(2.1)}). Note the range of r.
Moreover, it should be noticed that, for $r \in [0,1]$,
the inequality in eq. (\ref{(4.9)}) is strongly violated.}
\label{f:casoB}
\end{center}
\end{figure}

\begin{figure}
\begin{center}
\begin{tabular}{ccc}
\includegraphics[width=7cm]{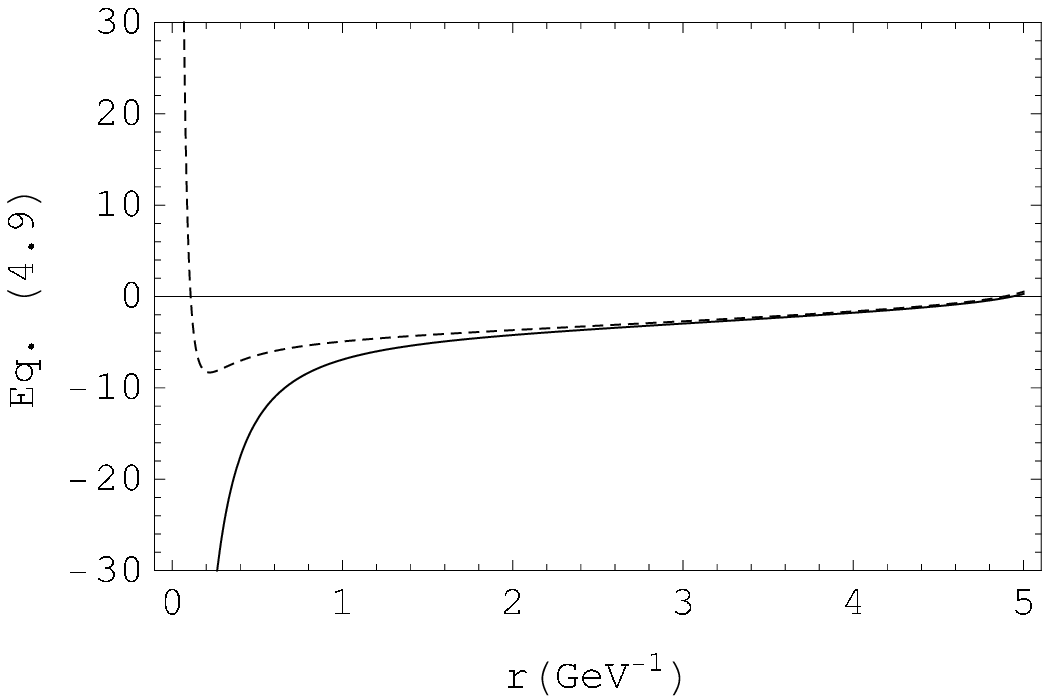} &
\hspace{2truecm}                                   &
\includegraphics[width=7cm]{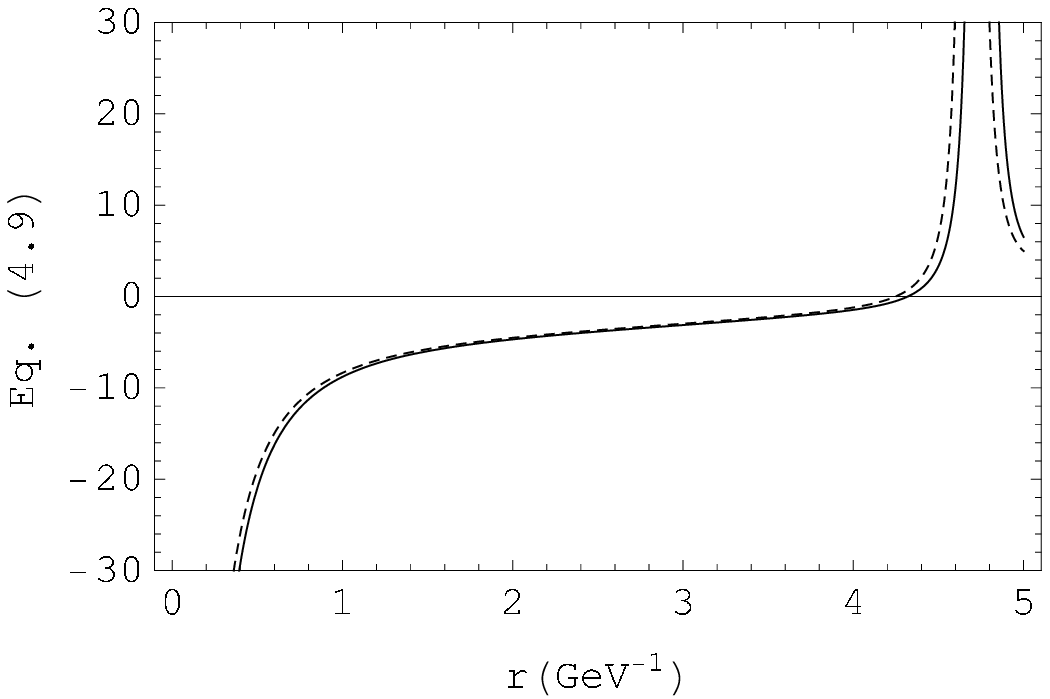} 
\end{tabular}
\caption{
The same as figure 1. Moreover, $b=-0.579$.}
\label{f:casoC}
\end{center}
\end{figure}

\begin{figure}
\begin{center}
\begin{tabular}{ccc}
\includegraphics[width=7cm]{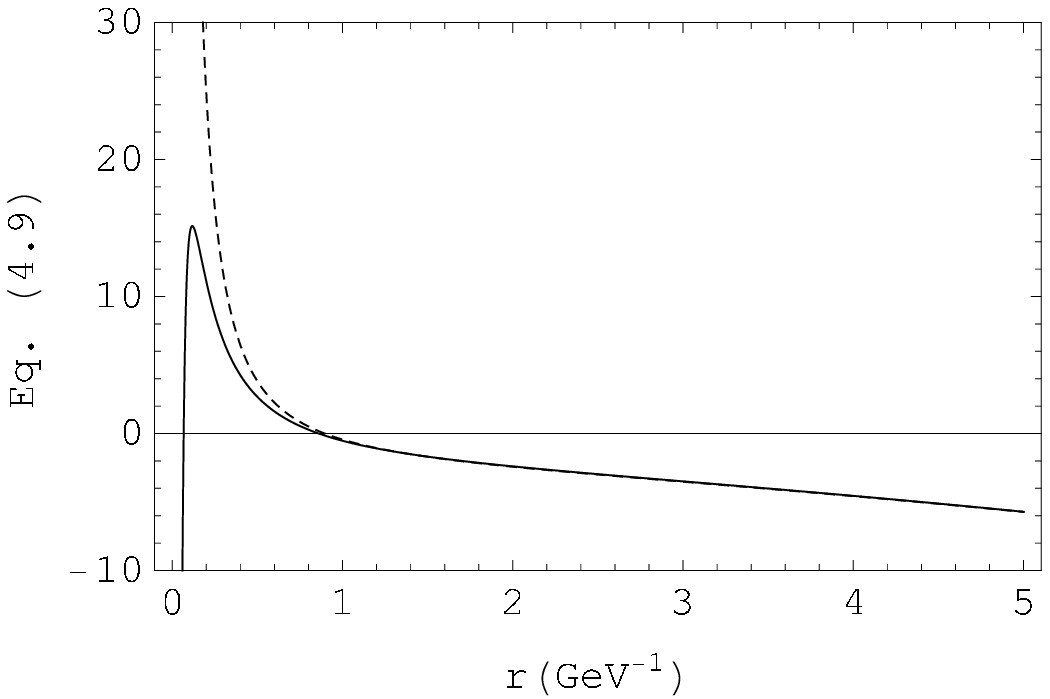} &
\hspace{2truecm}                                    &
\includegraphics[width=7cm]{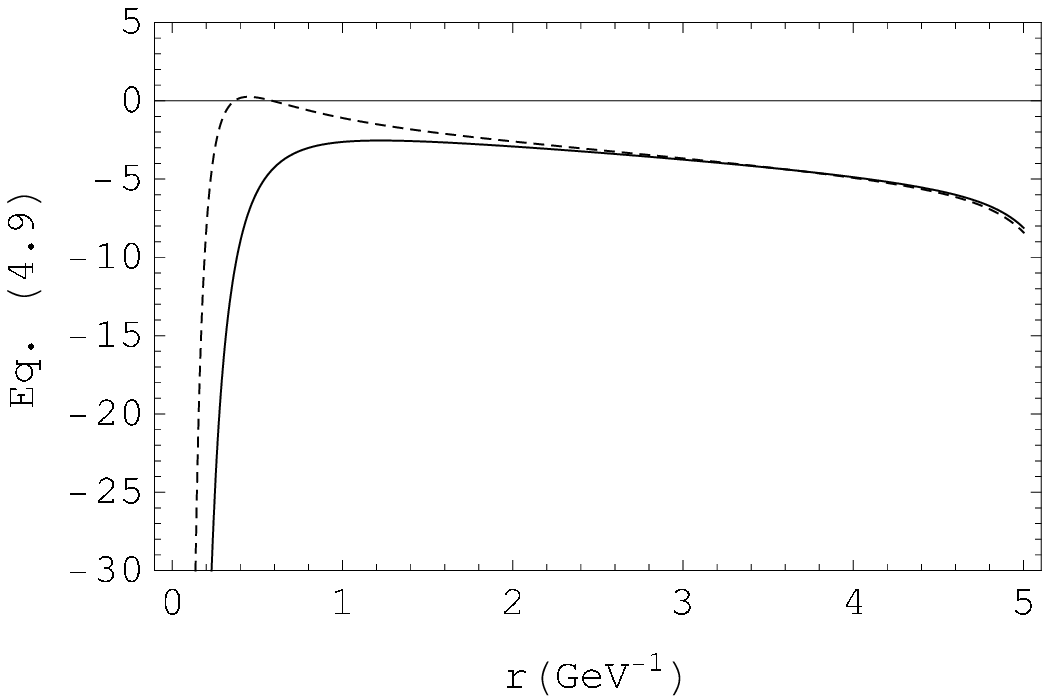}
\end{tabular}
\caption{
The same as figure 3. Here $Q^2_{f,g}$ are chosen as in
section \ref{s:casoD}.}
\label{f:casoD}
\end{center}
\end{figure}

In figure \ref{f:casoA} we have plotted the
left-hand side of eq. (\ref{(4.9)}) for the
$(R, Q^2) \equiv (R_f,Q^2_f)$  (left panel) and
$(R, Q^2) \equiv (R_g,Q^2_g)$ (right panel). The
light quark mass, $m=0.300\,\,GeV$, $a=0.308\,\, GeV^{2}$ and
$E = 1.9\,\,GeV$ in $R_f$ and $R_g$
(cfr eqs. (\ref{(3.2)})-(\ref{(3.3)})).
The plots in figures \ref{f:casoB}-\ref{f:casoD} are obtained by using
the values collected in their captions.
It should be observed that in figure \ref{f:casoA} we have used a
confining linear potential and for $Q^2$ the choice in section \ref{s:casoA}.
The inequality in (\ref{(4.9)}) is satisfied
for almost the whole physical range of $r$.

In figure \ref{f:casoB} the logarithmic potential has been
considered (cf. section \ref{s:casoB})
with $r_0=1\,\,\mathrm{GeV}^{-1}$. Also in this case we
do not have direct phenomenological information on the
values of the parameters. Smaller values for $r_0$ are responsible
for the violation of the inequality (\ref{(4.9)}).

In figures \ref{f:casoC} and \ref{f:casoD} the Cornell potential
is considered. In these figures the values of the parameters are taken,
as already said, from the phenomenological analysis.
In fig. \ref{f:casoC} the inequality is violated for $Q^2_g$
in the whole range of $r$. While the case
inspired by ordinary quantum mechanics (cf fig. \ref{f:casoD})
violates the inequality in the region of small $r$.

\section{Stokes and anti-Stokes lines}
\label{s:6}

In the application of the phase-integral method to Eq. (1.1), a
concept of particular relevance is the one of Stokes and anti-Stokes
lines. By definition, the differential $dw=q(z)dz$ (see (1.4)) is purely
imaginary along a Stokes line, and real along an anti-Stokes line.
Thus, the Stokes lines are lines along which the absolute value of
$e^{i w(z)}$ increases or decreases most rapidly, while the anti-Stokes
lines are level lines for constant absolute values of $e^{i w(z)}$
\cite{Froman02}.

For example, for the case studied in our subsection 3.D one can evaluate
at complex $r=A e^{i \theta}$ the phase integral (3.12). One then finds,
after repeated application of the Gauss representation of complex
numbers, and upon defining
\begin{equation}
{\widetilde A} \equiv 1+{A \over 2\eta}\cos(\theta), \;
{\widetilde B} \equiv {A\over 2\eta}\sin(\theta),
\label{(6.1)}
\end{equation}
\begin{equation}
{\widetilde \theta} \equiv {\rm arctan} \left(
{A \sin (\theta) \over (2\eta + A \cos (\theta))} \right),
\label{(6.2)}
\end{equation}
\begin{equation}
\alpha \equiv \sqrt{A \over 2\eta} \cos {\theta \over 2}
+\Bigr({\widetilde A}^{2}+{\widetilde B}^{2}\Bigr)^{{1\over 4}}
\cos {{\widetilde \theta}\over 2},
\label{(6.3)}
\end{equation}
\begin{equation}
\beta \equiv \sqrt{A \over 2\eta} \sin {\theta \over 2}
+\Bigr({\widetilde A}^{2}+{\widetilde B}^{2}\Bigr)^{{1\over 4}}
\sin {{\widetilde \theta}\over 2},
\label{(6.4)}
\end{equation}
\begin{equation}
\varphi \equiv {\rm arctan}{\left[
\sqrt{A \over 2\eta}\sin {\theta \over 2}
+\Bigr({\widetilde A}^{2}+{\widetilde B}^{2}\Bigr)^{{1\over 4}}
\sin {{\widetilde \theta}\over 2} \right] \over
\left[\sqrt{A \over 2\eta}\cos {\theta \over 2}
+\Bigr({\widetilde A}^{2}+{\widetilde B}^{2}\Bigr)^{{1\over 4}}
\cos {{\widetilde \theta}\over 2} \right]},
\label{(6.5)}
\end{equation}
the following split of $w_{f}(r)$ into real and imaginary part:
\begin{equation}
{\rm Re} \; w_{f}=2\eta \left[\sqrt{A \over 2\eta}
\Bigr({\widetilde A}^{2}+{\widetilde B}^{2}\Bigr)^{{1\over 4}}
\cos \left({\theta-{\widetilde \theta} \over 2}\right)
+{1\over 2}\log(\alpha^{2}+\beta^{2})\right],
\label{(6.6)}
\end{equation}
 \begin{equation}
{\rm Im} \; w_{f}=2\eta \left[\sqrt{A \over 2\eta}
\Bigr({\widetilde A}^{2}+{\widetilde B}^{2}\Bigr)^{{1\over 4}}
\sin \left({\theta+{\widetilde \theta} \over 2}\right)
+\varphi \right].
\label{(6.7)}
\end{equation}
From what we said before, along an anti-Stokes line, $dw_{f}$
is real, and hence ${\rm Im}w_{f}$ is constant.
We thus find from Eqs. (6.5) and (6.7) the transcendental equation
\begin{equation}
{\rm arctan}{\left[
\sqrt{A \over 2\eta}\sin {\theta \over 2}
+\Bigr({\widetilde A}^{2}+{\widetilde B}^{2}\Bigr)^{{1\over 4}}
\sin {{\widetilde \theta}\over 2} \right] \over
\left[\sqrt{A \over 2\eta}\cos {\theta \over 2}
+\Bigr({\widetilde A}^{2}+{\widetilde B}^{2}\Bigr)^{{1\over 4}}
\cos {{\widetilde \theta}\over 2} \right]}
+\sqrt{A \over 2 \eta}\Bigr({\widetilde A}^{2}
+{\widetilde B}^{2}\Bigr)^{{1\over 4}}
\sin \left({\theta+{\widetilde \theta} \over 2}\right)=
{\rm const}.
\label{(6.8)}
\end{equation}
Moreover, since $dw_{f}$ is purely imaginary along a Stokes line,
we are led to consider the equation
$$
{\rm Re} \; w_{f}={\rm constant}.
$$
This becomes, from (6.6), the transcendental equation
\begin{equation}
\left[\sqrt{2A \over \eta}
\Bigr({\widetilde A}^{2}+{\widetilde B}^{2}\Bigr)^{{1\over 4}}
\cos \left({\theta-{\widetilde \theta} \over 2}\right)
+\log(\alpha^{2}+\beta^{2})\right]= {\rm const}.
\label{(6.9)}
\end{equation}

\begin{figure}
\begin{center}
\includegraphics[width=9cm]{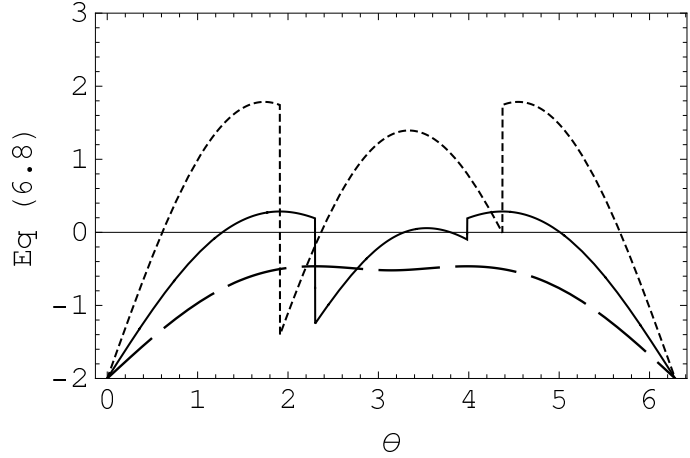}
\caption{Here we plot the left-hand side of eq. (\ref{(6.8)}) minus 2, which
is the value of the constant in the same equation, {\it versus} $\theta$.
The curves have been obtained for $\eta=(0.5,1,2)$
(dashed, continuous, long dashed) lines (cf eq. (\ref{(3.10)})) and $A=3$.
This figure shows that solutions to eq. (\ref{(6.8)}) exist.}
\label{(fig1)}
\end{center}
\end{figure}

\begin{figure}
\begin{center}
\includegraphics[width=9cm]{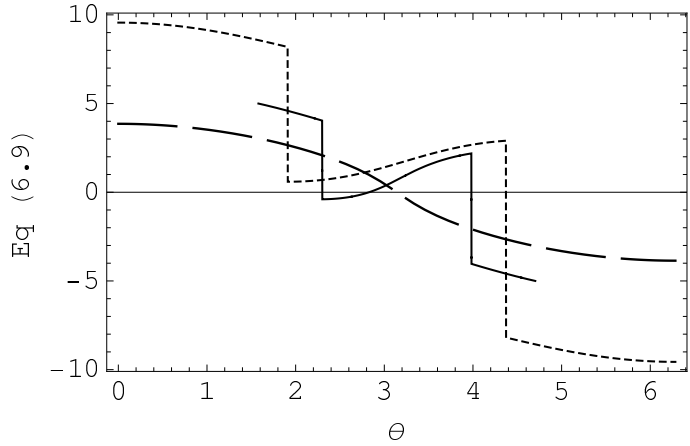}
\caption{Here we plot the left-hand side of eq. (\ref{(6.9)})
{\it versus} $\theta$.
The plot is obtained for $\eta=(0.5,1,2)$ (dashed, continuous,
long dashed) lines (cf eq. (\ref{(3.10)}))  and $A=3$.
As in the case of fig. \ref{(fig1)}, solutions to eq. (\ref{(6.9)}) exist.}
\label{(fig2)}
\end{center}
\end{figure}

\begin{figure}
\begin{center}
\includegraphics[width=9cm]{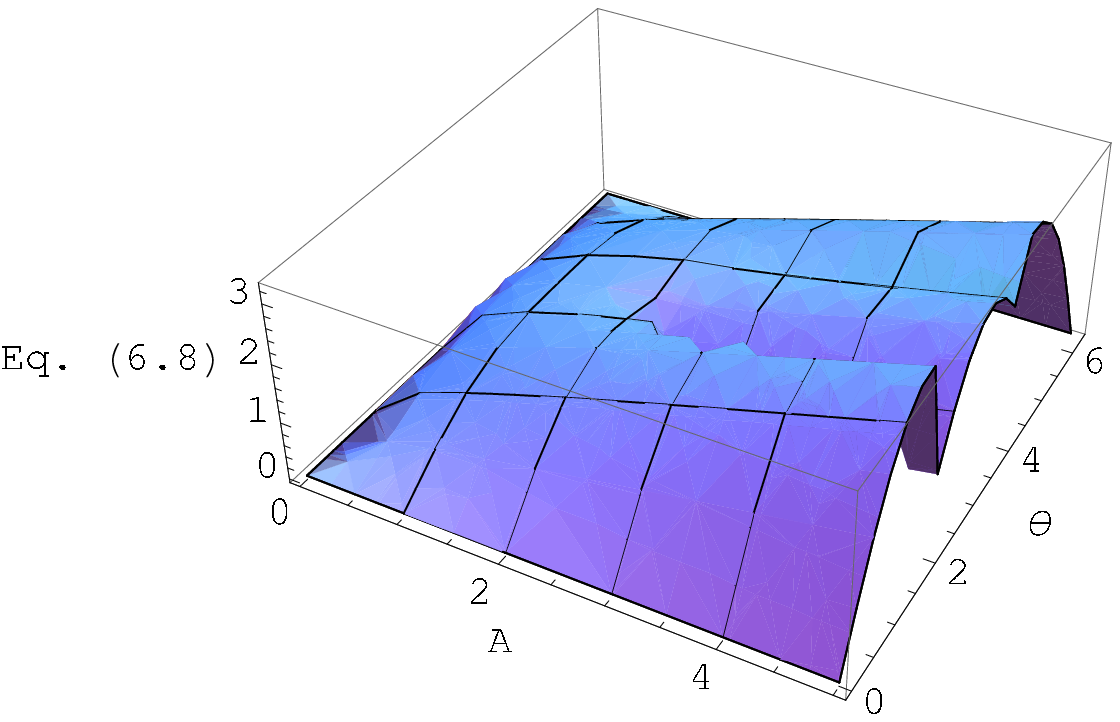}
\caption{We plot the left-hand side of eq. (\ref{(6.8)}) {\it versus}
$\theta$ and $A$ for $\eta = 2$ GeV.
}
\label{(fig3)}
\end{center}
\end{figure}

\noindent
In general, we cannot give analytical solutions to the
equations (\ref{(6.8)}) and (\ref{(6.9)}).
However, the fact that, for reasonable values of the parameters,
solutions to such equations exist is crucial.
In this respect, in figs. \ref{(fig1)} and \ref{(fig2)} we show that,
for $\eta=(0.5,1,2)$ and $A=3$, they can be solved
for a constant value and for zero, respectively. In particular,
eq.  (\ref{(6.8)}) has either zero or six roots depending on the choice
of the value for the constant, unlike the case of eq. (\ref{(6.9)}),
where at most three zeros can be found depending on the constant.

\begin{figure}
\begin{center}
\begin{tabular}{ccc}
\includegraphics[width=7cm]{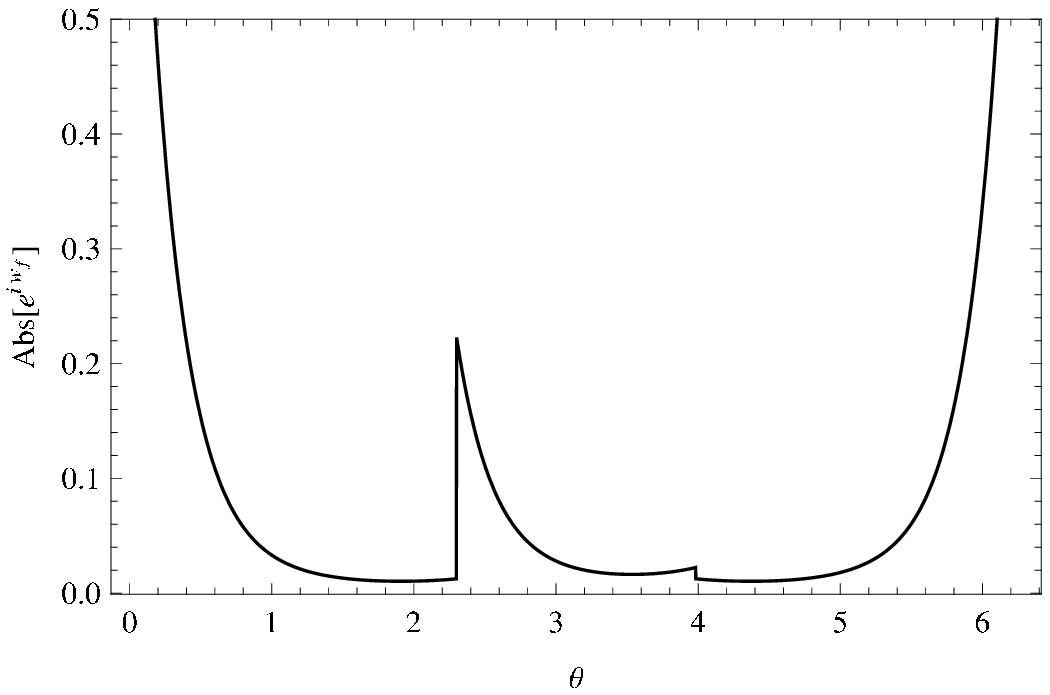} &
\hspace{2truecm}                                   &
\includegraphics[width=7cm]{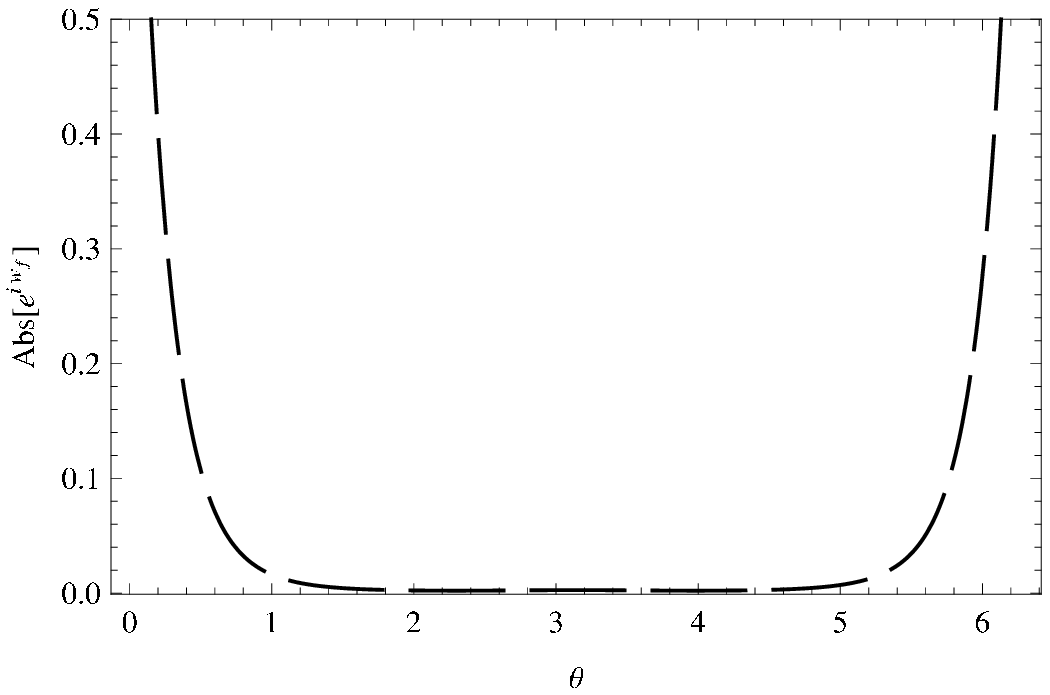} 
\end{tabular}
\caption{Here we plot $|e^{i w_f}|$ {\it versus} $\theta$.
The curves have been obtained for $\eta=(1,2)$
(continuous, long dashed) lines (cf eq. (\ref{(3.10)})) and $A=3$.}
\label{f:eiwf}
\end{center}
\end{figure}

Following what we say at the beginning of this section, 
the absolute value of $e^{iw_f}$ increases of decreases
along the Stokes lines while it remains constant along 
anti-Stokes lines. Figure \ref{f:eiwf}  displays this behaviour
in a neat way. 

\section{Concluding remarks and open problems}

Second-order equations for relativistic systems have been investigated
along many years, including the work in \cite{Goldberg}, and 
supersymmetric extensions considered in \cite{Cooper}. In ordinary
quantum mechanics, the most powerful choice of base function is
the Langer choice \cite{Langer, Crothers, Linnaeus}, but the peculiar
technical difficulties of the effective potentials (2.2) and (2.3)
for the Dirac equation cannot be solved in the same way, and one has
rather to resort to the JWKB method along the lines in \cite{Lazur}.
It was here our intention to investigate potentialities and limits
of the phase-integral method, which actually differs 
from JWKB methods \cite{Froman02}. Our results are of qualitative
nature, while we fail to obtain bound-state energies from the 
integrals in sections 2 and 3.
At a deeper level, the problem arises of solving coupled systems of
first-order ordinary differential equations which, when decoupled, give
rise to a pair of equations of the form (1.1). The phase-integral
method, originally developed for second-order equations of the form (1.1),
should have implications for the solutions of the original first-order
system as well. This expectation should be made precise, and its relation
with the JWKB method should be elucidated.

Although the decoupled second-order equations obtained from the
radial Dirac equation are formally analogous to the second-order
equations to which the phase-integral method can be applied, the actual
implementation is much harder because the ``potential'' terms
$R_{f}$ and $R_{g}$ therein contain complicated denominators built from
the potentials $V_{S}$ and $V_{V}$ in the
radial Dirac equation \cite{PHSTB-77-065005, JPAGB-32-5643}.
This implies that the actual choice of base function $Q$ is a difficult
problem. In Sec. III we have described some possible choices of $Q$,
and in Sec. IV we have arrived at the majorization (4.9) to select $Q$,
tested numerically in Sec. V. The analysis  of (4.9) for the
Cornell potential  shows that an appropriate basis function can be found for  
the case $k=-1$ (see fig. 3). Moreover, for the logarithmic potential 
the plots displayed in fig. 2 show that (4.9) is not fulfilled 
in the whole range of $r$.
The investigation of Stokes and anti-Stokes lines in Sec. VI 
is also, as far as we know, original in
our context. It remains to be seen, however, whether such lines 
can be of direct  phenomenological interest.

The work in Ref. \cite{PHSTB-77-065005}, despite being devoted to the
amplitude-phase method, did not investigate our same technical issues.
Thus, no obvious comparison can be made. The years to come will hopefully
tell us whether choices of $Q$ satisfying (4.9) exist for which the
$F$-matrix in (2.18) can be actually evaluated. In the affirmative case,
one would gain conclusive evidence in favour of the superiority of the
phase-integral method. In the negative case, one would instead gain
a better understanding of the boundaries to our knowledge.

\acknowledgments
G. Esposito is grateful to the Dipartimento di Scienze Fisiche of
Federico II University, Naples, for hospitality and support, and to
A.Yu. Kamenshchik for stimulating his interest in the theory of
Stokes lines.

\end{document}